\documentclass[twocolumn,prl,amsmath,amssymb]{revtex4}

\usepackage{graphicx}% Include figure files
\usepackage{dcolumn}% Align table columns on decimal point
\usepackage{bm}% bold math
\newcolumntype{.}{D{.}{.}{-1}}

\begin{document}

\preprint{APS/123-QED}

\title{$^{87}$Sr lattice clock with inaccuracy below 10$^{-15}$}

\author{Martin M. Boyd, Andrew D. Ludlow, Sebastian Blatt, Seth M. Foreman, Tetsuya Ido$^{\dag}$, Tanya Zelevinsky, and Jun Ye }
 \affiliation{JILA, National Institute of Standards and Technology and University of Colorado,
  Department of Physics, University of Colorado, Boulder, CO 80309-0440}

\date{\today}

\begin{abstract}
Aided by ultra-high resolution spectroscopy, the overall systematic
uncertainty of the $^{1}S_{0}$-$^{3}P_{0}$ clock resonance for
lattice-confined $^{87}$Sr has been characterized to
$9\times10^{-16}$. This uncertainty is at a level similar to the
Cs-fountain primary standard, while the potential stability for the
lattice clocks exceeds that of Cs. The absolute frequency of the
clock transition has been measured to be 429,228,004,229,874.0(1.1)
Hz, where the $2.5\times10^{-15}$ fractional uncertainty represents
the most accurate measurement of a neutral-atom-based optical
transition frequency to date.

\end{abstract}


\maketitle

The significant advances in femtosecond comb technology
\cite{Diddams1,Udem, Ye1} in the past decade have sparked immense
interest in atomic clocks based on optical transitions
\cite{Diddams2}. These transitions have large line quality factors
($Q$) \cite{Boyd1, Bergquist1}, which will provide orders of
magnitude improvement in clock stability over state-of-the-art
microwave clocks. An optical clock based on a single trapped Hg$^+$
ion has recently surpassed Cs fountain clocks \cite{SyrteCs, NISTCs}
in terms of accuracy, with clock systematics reduced to
$7\times10^{-17}$ \cite{Bergquist2}.  Other high accuracy ion
standards include Sr$^+$ \cite{NPLScience,NRC} and Yb${^+}$
\cite{PTBYb}.  The high line $Q$ allows a stability comparable to
the best achieved thus far with Cs, despite the fact that the
single-ion signal-to-noise ratio ($S/N$) is drastically reduced
compared to microwave systems which typically employ $\sim$$10^5$
atoms. Optical lattice clocks show promise for reaching a level of
accuracy comparable to the ion systems, with significantly improved
stability due to the large number of atoms involved in the
measurement.  This stability gain has spurred an intensive
investigation of lattice clocks based on spin-forbidden transitions
in alkaline-earth atoms, specifically in Sr \cite{Boyd1, Katori1,
KatoriNature, Ludlow1, LeTargat1} and Yb \cite{Fortson1,NistYb},
where the trapping potential is designed to allow accurate
measurements effectively free of both ac Stark shifts \cite{Katori1,
Katori2, Anders1} and motional effects which can hamper optical
clocks based on atoms in free space \cite{SterrNew, OatesNew, Ido1}.

\begin{figure}[t]
\resizebox{8.5cm}{!}{
\includegraphics[angle=0]{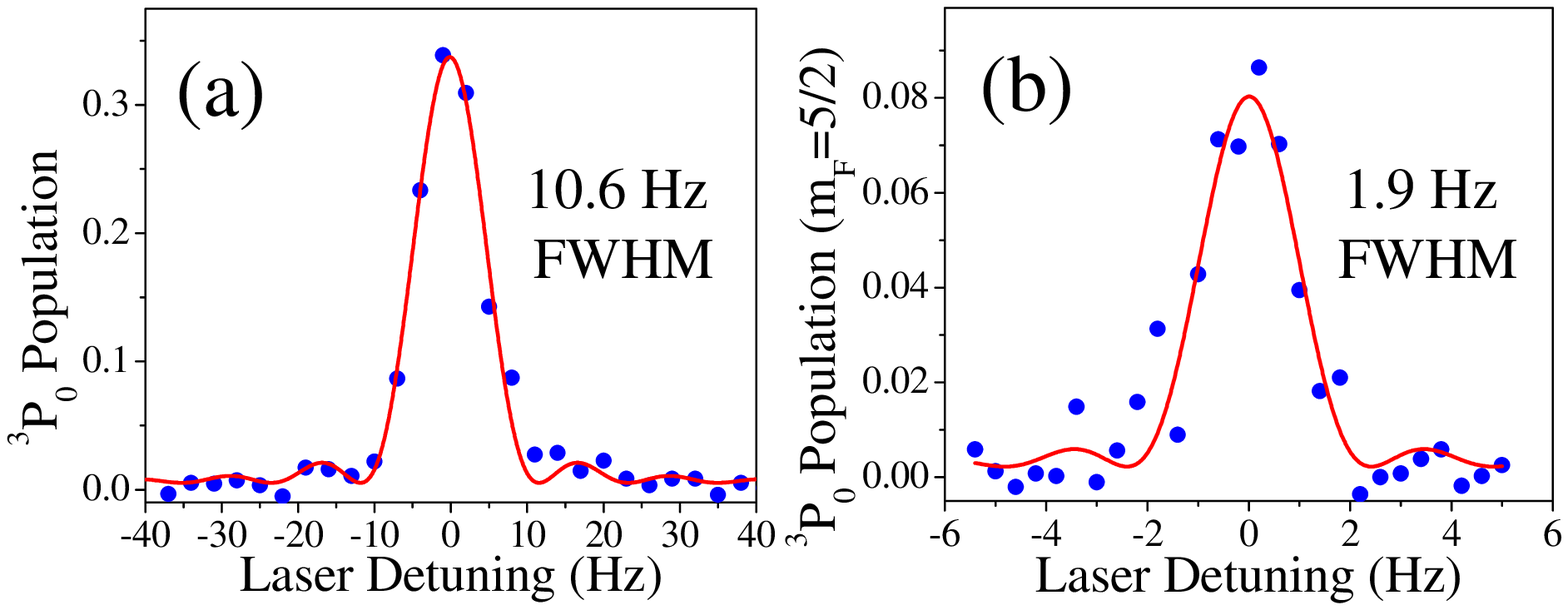}}
\caption{\label{Fig1}(color online) High resolution spectroscopy of
the $^1S_0$-$^3P_0$ transition.  (a) A Fourier-limited resonance
profile for typical operating parameters in the experiment. A
sinc$^2$ fit is shown in red, giving a linewidth (FWHM) of 10.6(3)
Hz ($Q$ $\simeq$ $4\times10^{13}$). (b) Spectroscopy of an isolated
nuclear spin component yields a linewidth of 1.9(2) Hz ($Q$ $ \simeq
2.3\times10^{14}$), consistent with the 1.8 Hz Fourier limit for the
0.48 s probe time. The spectra in both (a) and (b) are taken without
averaging or normalization, and the vertical axes are scaled by the
total number of atoms ($2\times10^4$). The $S/N$ in (a) is limited
by shot to shot atom number fluctuation, whereas the probe laser
frequency noise is the dominant effect in (b).}
\end{figure}

While the clock-stability benefits provided by the optical lattice
method are now clear \cite{NistYb, Boyd1}, reaching the accuracy
level of the microwave standards remains a paramount issue in the
field. Recently, great strides towards this goal have been taken, as
a troublesome 4$\sigma$ disagreement between the first two high
accuracy experiments using $^{87}$Sr \cite{KatoriNature, Ludlow1}
has been resolved by a third independent investigation
\cite{LeTargat1}, and a revised report by the authors of Ref.
\cite{KatoriNature} published shortly thereafter \cite{KatoriJSP}.
Agreement between the three groups speaks strongly for the lattice
clock as a future candidate for redefinition of the SI second;
however, to be competitive with the current Cs fountain clocks the
overall systematics must be reduced well below the $10^{-15}$ level.

In this Letter, we report a detailed study of the systematic
uncertainty associated with the $^{87}$Sr $^1S_0$-$^3P_0$ clock
transition frequency at the level of $9\times10^{-16}$.  This
measurement, aided mainly by the record level line $Q$ achieved
recently \cite{Boyd1}, shows that the Sr lattice clock can reach an
accuracy level competitive with Cs fountains, while the potential
stability for the system is far greater.  An absolute frequency
measurement of the transition is also reported with an uncertainty
of $2.5\times10^{-15}$, limited by a Cs-calibrated NIST H-maser
reference.

\begin{figure}[t]
\resizebox{8.5cm}{!}{
\includegraphics[angle=0]{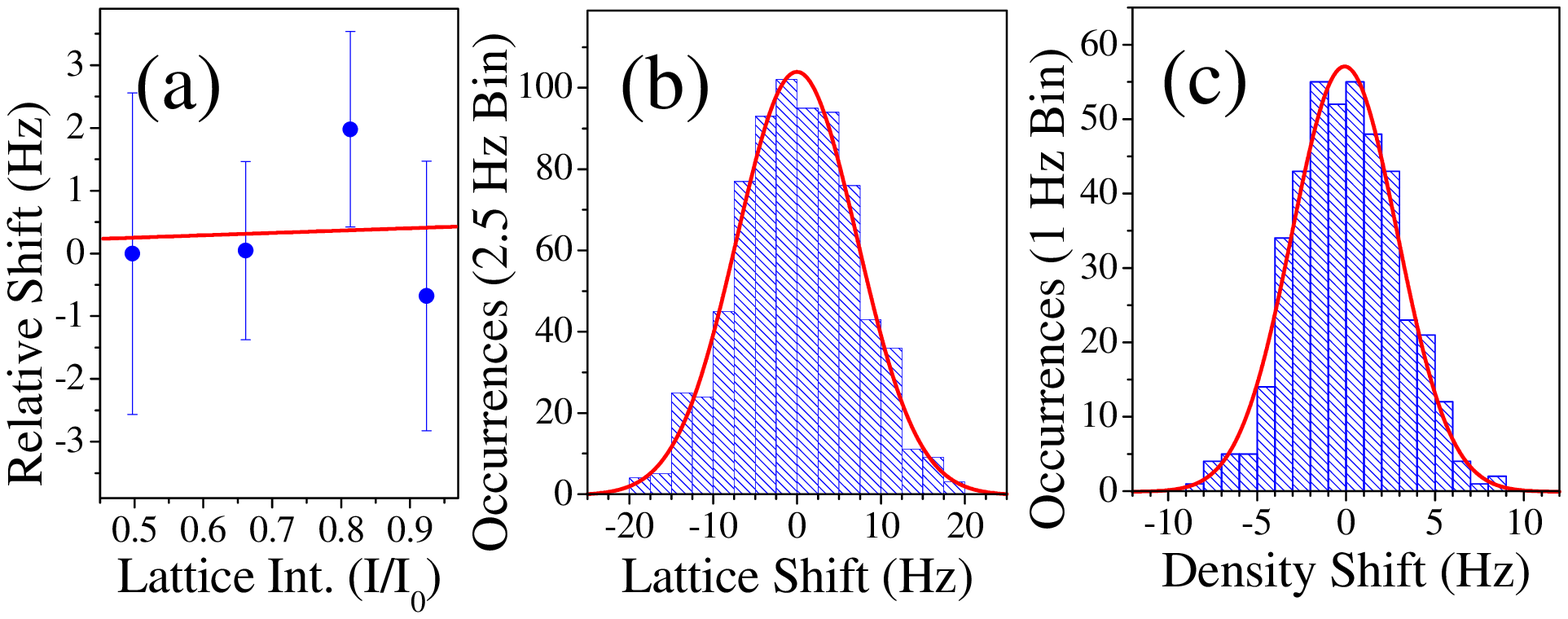}}
\caption{\label{Fig3}(color online) (a) Measurement of the lattice
Stark shift achieved using four interleaved intensities during a
single scan of the clock transition. For this single measurement the
shift is 0.4(4.4) Hz/$I_0$. (b) Histogram of 776 interleaved
measurements (such as (a)) of the Stark shift for our typical
intensity $I_0$. (c) Histogram of 422 measurements in which the
atomic density is varied by a factor of 5 during spectroscopy.
Gaussian fits to the data in (b) and (c) are shown in red.}
\end{figure}

Full details of the cooling and trapping system used in this work
are discussed elsewhere \cite{Loftus1, Ludlow1}. In brief, $^{87}$Sr
atoms are captured from a thermal beam into a magneto-optical trap
(MOT) based on the $^1S_0$-$^1P_1$ cycling transition.  Second stage
cooling, using a dual frequency $^1S_0$-$^3P_1$ MOT
\cite{Mukaiyama1}, is performed concurrently with the loading of a
vertical one-dimensional lattice, yielding $\sim$2 $\times$ $10^4$
atoms at a temperature of 1.5 $\mu$K.  The lattice is operated at
the Stark cancelation wavelength \cite{Anders1} with an intensity
$I_0$ = 5 kW/cm$^2$ (83$\%$ of which forms the standing wave due to
window losses), resulting in measured longitudinal and radial trap
frequencies of 40 kHz and 125 Hz respectively.  The atoms are
distributed over $\sim$80 lattice sites with a density $\rho_0$ =
$5\times10^{11}$ cm$^{-3}$. The spectroscopy sequence for the 1 mHz
$^1S_0$-$^3P_0$ clock transition begins with a Rabi pulse from a
highly stabilized diode laser \cite{Ludlow2} that is co-propagated
and co-polarized with the lattice laser.  With some atoms shelved in
the $^3P_0$ state, the remaining $^1S_0$ population is removed from
the lattice by exciting the $^1S_0$-$^1P_1$ transition. The $^3P_0$
atoms are then driven back to the ground state, by pumping through
intermediate states, and the population is measured by again driving
the $^1S_0$-$^1P_1$ transition and detecting scattered photons. This
process is repeated each time atoms are loaded into the lattice, as
the laser frequency is tuned. The time window for the Rabi pulse is
varied within 40-480 ms depending on the desired Fourier width
(22-1.8 Hz).

The $^1S_0$ ($F$=9/2) - $^3P_0$ ($F$=9/2) transition, facilitated by
nuclear-spin induced state mixing \cite{Kluge1}, suffers from a
differential Land\'e $g$-factor between the clock states, with the
$^3P_0$ sensitivity being $\sim$60$\%$ larger than that of the
ground state. The resultant Zeeman shift of -109 Hz/(G $m_F$)
\cite{Boyd1} (1 G = $10^{-4}$ T) can be a limitation in terms of the
achievable accuracy and line $Q$ in the presence of magnetic fields.
Figure 1(a) shows a spectrum for a 80 ms probe time, representing
the parameters typically used in the work reported here, yielding a
FWHM (full width at half maximum) linewidth of 10.6(3) Hz.  The
spectrum shown here supports a clock instability of less than
$3\times10^{-15}$ at 1 s.  For atom-shot-noise limited spectra of
the same width, and reasonable improvements to the duty cycle, the
number could be reduced by more than an order of magnitude. The
narrowest resonances have so far been achieved when a resolved
nuclear sublevel is used for spectroscopy as shown in Fig. 1(b).
Here, linewidth limitations from magnetic fields or state-dependent
Stark shifts are eliminated, and widths below 2 Hz are repeatably
observed.

As a general approach for evaluating systematics, an interleaved
scheme is used where the parameter of interest is cycled through
different values, synchronized with each frequency step of the probe
laser across the resonance.  The interleaved data is then separated
into resonance profiles for each parameter value, allowing the
center frequency (relative to the laser cavity), and more
importantly the slope of the frequency shift, to be measured for a
variety of system parameters in a short time. This method allows us
to measure shifts against the probe laser, which has a stability
superior to our available microwave reference \cite{Ludlow2}.

Of the many effects to be characterized for an optical lattice
clock, the effect of the lattice laser itself remains a focal point.
The differential light shifts of the clock states due to the scalar,
vector, and tensor polarizabilities all vary linearly with trap
intensity and can be strongly suppressed with an appropriate choice
of lattice wavelength \cite{Katori2,Fortson1}. Higher-order Stark
shifts, due to the hyperpolarizability of the clock states, are
negligibly small ($<10^{-17}$)\cite{Anders1} at our operating
intensity and wavelength. Hence, a linear extrapolation to the
zero-intensity clock frequency is sufficient to characterize the
total Stark shift from all contributors mentioned above. An example
of this is shown in Fig. 2(a) where four different values of the
lattice intensity are interleaved during a trace taking less than
one minute. Using 2, 4, or 8 lattice intensity values, 776
interleaved measurements revealed that for a wavelength $\lambda_0$
= 813.4280(5) nm, the Stark shift is -108(257) mHz/$I_0$. A summary
of the lattice Stark shift measurements is shown in Fig 2(b) as a
histogram, along with a gaussian fit of the data.

The effect of atomic density on the transition frequency is explored
in a similar fashion as densities ranging within (0.2-1)$\rho_0$ are
interleaved (by varying the number of atoms in the MOT). A histogram
of 422 measurements of the density effect is shown in Fig. 2(c),
resulting in a shift coefficient of 3(140) mHz/$\rho_0$. Notably the
upper limit of the density-related fractional frequency shift of
$5.6\times10^{-28}$ cm$^{-3}$ is $\sim$$10^6$ times smaller than for
Cs \cite{NISTCs, SyrteCs}.

The ten nuclear-spin sublevels of the clock transition result in
systematic effects related to magnetic and optical fields. For
example, the asymmetric distribution of population among the
sublevels can be a central systematic issue when using unpolarized
atomic samples, as any $m_F$-dependent magnetic or optical
interaction can cause a frequency shift, even if the sub-levels are
shifted symmetrically about the center. The differential $g$-factor
of the clock states provides the most significant effect as it leads
to a sensitivity to magnetic fields of nearly 500 Hz/G for the
stretched states. Three orthogonal sets of Helmholtz coils are used
to characterize frequency shifts caused by the Zeeman sensitivity of
the nuclear-spin sublevels. Figure 3 summarizes the characterization
of magnetic field effects along one of these three axes. For each
direction, the transition linewidth is used to find the field
minimum as shown in Fig. 3(a). The narrow 10 Hz resonances allow the
field zero to be constrained within 10 mG for each axis. Frequency
shift sensitivity is explored using the interleaved scheme with the
results for the featured axis shown in Fig. 3(b). Here the average
values for 112 measurements are shown, yielding a slope of 26(4)
Hz/G.  Similar measurements were performed for the other two axes
yielding 22(7) Hz/G and 12(3) Hz/G.  The fields for the three axes
are zeroed below 5, 5, and 10 mG respectively, resulting in a total
Zeeman uncertainty of $5.3\times10^{-16}$. This gives insight into
the minimal effect of the vector light shift which causes symmetric
$m_F$-dependent shifts proportional to the degree of lattice
ellipticity and trapping intensity \cite{Fortson1, Katori2}. The
resultant splitting for the stretched states is estimated as less
than 8 (Hz/rad)/$I_0$. To combat this effect, a high extinction
polarizer ($>$$10^{4}$) is used for the lattice and probe beam, and
while the vacuum chamber windows likely reduce the purity of the
linear polarization, rotations of even a few degrees are equivalent
to a sub mG residual field, attesting to the insignificance of this
effect compared to the differential $g$-factor.

\begin{figure}[t]
\resizebox{8.5cm}{!}{
\includegraphics[angle=0]{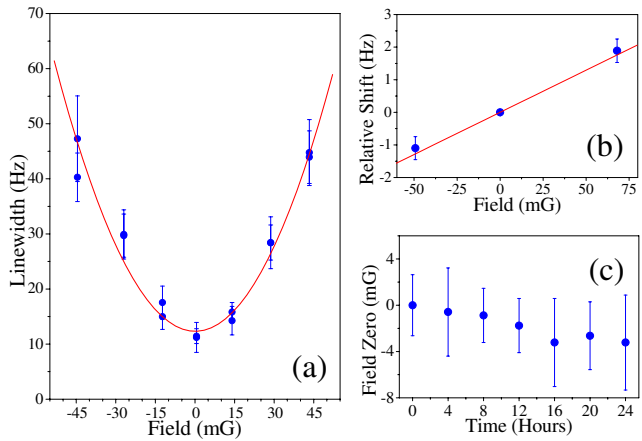}}
\caption{\label{Fig2}(color online) Effect of a magnetic field on
the transition linewidth and frequency. (a) The magnetic field is
calibrated using the width of the narrow resonance.  The data is fit
to a parabola, determining the field zero within 4 mG. (b) The
result of 112 interleaved measurements where the field is varied
during spectroscopy. The frequencies of the zero field values are
used as a reference for presentation purposes. The slope of all the
measurements for this axis yields an average value of 26(4) Hz/G.
(c) Summary of field calibrations during the 24-hour absolute
frequency measurement.}
\end{figure}

Systematics related to the probe laser were considered in two
respects. First, the probe can cause Stark shifts of the clock
states by coupling to external levels.  Second, asymmetric motional
sidebands could cause line pulling. This effect is minimal as the
sidebands are well resolved (even the radial sidebands are detuned
by more than ten times the transition width) and are only observed
for large probe intensities. These effects were checked
experimentally by varying the probe power by more than an order of
magnitude during 77 measurements.  To eliminate Stark shifts from
other sources, all lasers used for cooling, trapping, and detection
are switched with both acousto-optic modulators and mechanical
shutters.  Shifts from black body radiation (BBR) are also
considered \cite{DerevBBR}, including the effect of a nearby heated
vacuum window.

Table I summarizes the dominant systematic uncertainties for
spectroscopy of the clock transition, reported in terms of
fractional frequency. A total uncertainty of $0.88\times10^{-15}$ is
achieved, representing the first experimental verification that the
lattice technique can reach inaccuracies below the $10^{-15}$ level,
comparable with Cs fountains.  The largest uncertainties are limited
by technical issues such as a small dynamic range on the lattice
intensity and sensitivity to stray magnetic fields. Future work
using isolated spin states should allow orders of magnitude
reduction in the nuclear-spin related shifts, while significant
reductions in the lattice shift uncertainty can be achieved using a
larger range of intensities. Spin-polarizing the atoms can also
minimize collision shifts via Fermi suppression. However, unless the
spin polarization is pure and all atoms are in a single motional
state of the trap (possible but not yet achieved in a lattice
clock), the collision shift must still be evaluated experimentally.

\begin{table}[t]
\caption{Strontium Lattice Spectroscopy Error Budget}
\begin{ruledtabular}
\begin{tabular}{l..}
\multicolumn{1}{l}{Contributor} & \multicolumn{1}{l}{Correction ($10^{-15}$)} & \multicolumn{1}{l}{Uncertainty ($10^{-15}$)} \\
\hline AC Stark (Lattice) & 0.25 & 0.60 \\ AC Stark (Probe) & -0.02
& 0.12 \\ AC Stark (BBR) & 5.44 & 0.16
\\ Zeeman Effect & 0 & 0.53 \\ Density Shift & -0.01 & 0.33 \\
\hline Total & 5.66 & 0.88 \\
\end{tabular}
\end{ruledtabular}
\end{table}

To measure the absolute frequency of the $^1S_0$ - $^3P_0$
transition, a Cs-fountain-calibrated H-maser is used to stabilize a
radio frequency synthesizer located at NIST.  The synthesizer
modulates the amplitude of a 1320 nm laser, which is transferred to
JILA via a $\sim$4 km fiber link \cite{ye03, Foreman06}. The
modulation frequency of 950 MHz is compared to the repetition rate
of a femtosecond frequency comb locked to the spectroscopy laser.
The maser and transfer system provide a 1 s instability of
$2.5\times10^{-13}$, and for the work reported here, the maser is
calibrated to $1.7\times10^{-15}$. Passive transfer using the fiber
link has been found to introduce frequency offsets as large as
$1\times10^{-14}$, specifically related to periodic stretching and
compressing of the fiber length owing to daily temperature
variations. To eliminate this effect, the fiber length is stabilized
using a fiber stretcher controlled by comparison of the local
microwave phase at NIST with that of modulated light reflected back
from JILA. While in-loop measurements show the frequency transfer is
stabilized to a few parts in $10^{17}$, we assign a conservative
uncertainty of $1\times10^{-16}$ to account for other potential
errors \cite{FrenchFiber}.  The reference synthesizer for the
transfer can also cause frequency errors \cite{Diddams3} as drifts
in the synthesizer's temperature result in fractional shifts at the
level of $4\times10^{-14}$ (K/Hour)$^{-1}$. For the measurements
reported here, the synthesizer is placed in a temperature stabilized
enclosure and the temperature inside and outside the enclosure is
monitored, resulting in a correction of $-1.7(7)\times10^{-15}$.

\begin{figure}[t]
\resizebox{8.5cm}{!}{
\includegraphics[angle=0]{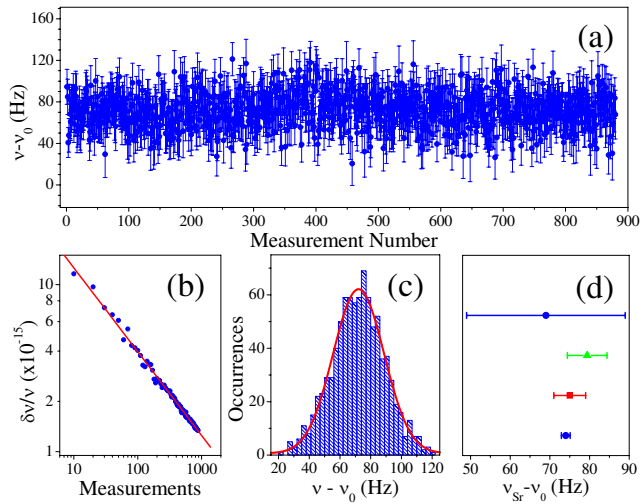}}
\caption{\label{Fig4}(color online) Absolute frequency measurement
of the $^1S_0$ - $^3P_0$ transition. (a) Counting record of 880
measurements taken over a 24 hour period (corrected for only the
maser offset in Table II). (b) The uncertainty averages down as
$N^{-0.501(4)}$, where $N$ is the number of measurements
(randomized), reaching 1.4$\times10^{-15}$ for the data set. (c) A
histogram of the frequency measurements in (a) with a gaussian fit
of the data shown in red. (d) Comparison of the final value,
$\nu_\mathrm{Sr}$, reported here with recent measurements by the
JILA (circle), SYRTE (triangle), and Tokyo (square) groups. Here the
offset frequency, $\nu_0$, is 429,228,004,229,800 Hz.}
\end{figure}

\begin{table}
\caption{Absolute Frequency Measurement Error Budget}
\begin{ruledtabular}
\begin{tabular}{l..}
\multicolumn{1}{l}{Contributor} & \multicolumn{1}{l}{Correction ($10^{-15}$)} & \multicolumn{1}{l}{Uncertainty ($10^{-15}$)} \\
\hline Sr Syst. (Table I) & 5.66 & 0.88 \\ Maser Calibration & -401.0 & 1.7 \\ Synth. Temp. Drift & -1.7 & 0.7 \\
Fiber Transfer & 0 & 0.1 \\
Gravitational shift & 1.25 & 0.02 \\
\hline Freq. Meas. Syst. & -395.8 & 2.0 \\
 Freq. Meas. Stat. & 0 & 1.4
\\\hline Total & -395.8 & 2.5 \\
 $\nu_\mathrm{Sr}-\nu_{0}$ & 74.0\,\mathrm{Hz} & 1.1\,\mathrm{Hz} \\
\end{tabular}
\end{ruledtabular}
\end{table}

A summary of 880 absolute frequency measurements spanning a full 24
hour period is shown in Fig. 4(a).  Each point corresponds to a 30
second measurement of an 11 Hz spectrum with a frequency uncertainty
of $\sim$20 Hz, consistent with the Allan deviation of the H-Maser.
The data averages down with gaussian statistics, as shown in Fig.
4(b) and in the histogram of Fig. 4(c). During the measurement, the
Sr chamber temperature was continuously monitored, and the magnetic
field was repeatedly calibrated (Fig. 3(c)) both by monitoring
transition linewidths and by employing the zeroing technique in Fig.
3(a). Table II summarizes the relevant corrections and uncertainties
associated with the absolute frequency measurement. The only
significant corrections not determined by direct frequency
measurements here are the BBR shift and the gravitational shift
arising from the difference in elevation of the NIST Cs fountain and
the JILA Sr lattice. The frequency of the $^1S_0$-$^3P_0$ transition
is 429,228,004,229,874.0(1.1) Hz, with the uncertainty mainly
limited by the maser calibration. Figure 4(d) shows that this value
agrees well with recent reports from the SYRTE \cite{LeTargat1} and
Tokyo \cite{KatoriJSP} groups as well as with our original value
\cite{Ludlow1}.  The final absolute frequency uncertainty of
2.5$\times10^{-15}$ corresponds to the most accurate optical
frequency measurement for neutral atoms to date, and falls short of
only the recent Hg$^+$ ion result \cite{Bergquist2} as the most
accurate optical measurement in any system.

We gratefully acknowledge technical contributions by S. Diddams and
T. Parker on maser transfer. We also thank D. Hudson and M. Ting for
help with the fiber link. This work was supported by ONR, NIST, and
NSF.

$\dag$ Current address for T. Ido: NICT, Tokyo, Japan.

\end{document}